\documentclass[draftclsnofoot,12pt, onecolumn]{IEEEtran}
\usepackage{amsmath}
\usepackage{epsfig}
\usepackage{color}
\usepackage[normalem]{ulem}
\usepackage{amssymb}
\usepackage{latexsym}
\usepackage{graphicx}
\usepackage{cite}

\title{Sensing of Unknown Signals over Weibull Fading Conditions}

\author{
Paschalis~C.~Sofotasios$^{1}$,~Mulugeta~K.~Fikadu$^{2}$,~Khuong~Ho-~Van$^{3}$ and~Mikko~Valkama$^{2}$ 
\\ 
\begin{normalsize} 
$^{1}$School of Electronic and Electrical Engineering, University of Leeds, LS$2$ $9$JT, Leeds, United Kingdom. 
\end{normalsize}
\\
\begin{normalsize} 
e-mail: $\rm p.sofotasios@\rm leeds.ac.uk$
\end{normalsize} 
\\
\begin{normalsize} 
$^{2}$Department of Communications Engineering, Tampere University of Technology, FI-33101, Tampere, Finland.
\end{normalsize}
\\
\begin{normalsize} 
e-mail: $\rm \left\lbrace mulugeta.fikadu; mikko.e.valkama \right\rbrace@\rm tut.fi$ 
\end{normalsize} 
 \\
\begin{normalsize} 
$^{3}$Department of Telecommunications Engineering, HoChiMinh City University of Technology, HoChiMinh City, Vietnam. 
\end{normalsize}
\\
\begin{normalsize} 
e-mail: $\rm  khuong.hovan@\rm yahoo.ca$
\end{normalsize} 
 }

\begin{document}

\maketitle

\begin{abstract}
Energy detection is a widely used method of spectrum sensing in cognitive radio and Radio Detection And Ranging (RADAR) systems. This paper is devoted to the analytical evaluation of the performance of an energy detector over Weibull fading channels. This is a flexible fading model that has been shown capable of providing accurate characterization of multipath fading in, e.g.,  typical cellular radio frequency range of 800${/}$900 MHz. A novel analytic expression  for the corresponding average probability of detection is derived in a simple algebraic representation which renders it convenient to handle both analytically and numerically. As expected, the performance of the detector is highly dependent upon the severity of fading  as even small variation of the fading parameters affect significantly the value of the average probability of detection. This appears to be  particularly the case in severe fading conditions. The offered results are  useful in evaluating the effect of multipath fading in energy detection-based cognitive radio communication systems and therefore they can be used in quantifying the associated trade-offs between sensing performance and energy efficiency in cognitive radio networks.  
 \end{abstract}

\section{Introduction}

\pubidadjcol

Detecting unknown signals is a critical topic in wireless communications and has attracted for years the attention of both Academia and Industry. The detection is typically realized in the form of spectrum sensing with energy detection (ED) constituting the most simple and popular method. The operating principle of ED is based on the deployment of a radiometer, which is a non-coherent detection device that measures the energy level of a received signal waveform over an observation time window. The obtained measure is subsequently compared with a pre-defined energy threshold which determines accordingly whether an unknown signal is present or not \cite{J:Haykin, B:Haykin, B:Bargava}.

Detection of unknown signals  over a flat band-limited Gaussian noise channel was firstly addressed by H. Urkowitz in \cite{J:Urkowitz}. There, he comprehensively derived analytic expressions for the corresponding  probability of detection, $P_{d}$,  and probability of false alarm, $P_{fa}$,  performance metrics. These measures are generally based on the statistical assumption that the decision statistics follow the central chi-square and the non-central chi-square distribution, respectively. A few decades later, this problem was revisited by Kostylev  \cite{C:Kostylev} who considered quasi-deterministic signals operating in fading conditions. 

Thanks to its low implementation complexity and no requirements for prior knowledge of the signal, the ED method has been widely associated with applications in RADAR systems while it has been also shown to be  particularly useful in emerging wireless technologies such as ultra-wideband communications and cognitive radio  \cite{C:Tellambura, C:Win}. In the former, the energy detector is exploited for borrowing an idle channel from authorized users, whereas in the latter it identifies the presence or absence of a deterministic signal and decides whether a primary licensed user is in active or idle state, respectively. According to the corresponding decision, the secondary unlicensed user either remains silent, or proceeds in utilizing the unoccupied band until the primary user becomes again active \cite{J:Haykin, C:Tellambura_2}. This opportunistic method has been  shown to  increase substantially the utilization of the already allocated radio spectrum, offering a considerable mitigation of the currently extensive spectrum scarcity \cite{J:Ghasemi, P:Paschalis_1, P:Paschalis_2, P:Paschalis_3, J:Theo, P:Paschalis_4, P:Paschalis_5}.

\pubidadjcol

It is recalled that fading effects create a detrimental impact on the performance of wireless communication systems \cite{Paschalis_1, Paschalis_2, Paschalis_3, Paschalis_4, Paschalis_5, Paschalis_6, Paschalis_7, Paschalis_8, Paschalis_9, Paschalis_10} and the references therein. 
Several studies have been devoted to the analysis of the performance of energy detection-based spectrum sensing for different communication and fading scenarios. Specifically, the authors in \cite{J:Alouini} derived closed-form expressions for the average probability of detection over Rayleigh, Rice and Nakagami-$m$ fading channels for both single-channel and multi-channel scenarios. Similarly, the ED performance in the case of equal gain combining over Nakagami${-}m$ multipath fading has been reported by \cite{C:Herath} whereas the performance in collaborative spectrum sensing and in relay-based cognitive radio networks has been investigated by \cite{C:Ghasemi, C:Sousa, J:Ghasemi-Sousa, C:Attapattu, J:Attapattu_2}. A semi-analytic method for analyzing the performance of energy detection of unknown deterministic signals was reported in \cite{J:Herath} and is based on the moment-generating function (MGF) method. This method was utilized in the case of maximal-ratio combining (MRC) in the presence of Rayleigh, Rice and Nakagami-$m$ fading in \cite{J:Herath} as well as  for the useful case of correlated Rayleigh and Rician fading channels in \cite{C:Beaulieu}. Finally, the detection of unknown signals in low signal-to-noise-ratio (SNR) over $K$-distributed ($K$), generalized $K$ ($K_{G}$) and the flexible $\eta{-}\mu$ and $\kappa{-}\mu$ generalized fading channels have been analyzed in \cite{C:Atapattu_3, J:Janti, J:Attapattu, C:Attapattu_2, J:Sofotasios}.

Weibull statistical distribution was proposed in 1937 and became widely known by Walodi Weibull in 1951. It has been used in several fields of natural sciences and engineering in applications relating to reliability, failure data analysis, weather forecasting and modelling of signal level dispersion in RADAR systems.  In the context of wireless communications, it has been employed in modelling  multipath waves propagating in non-homogeneous environments. Weibull
distribution is very flexible and has been shown to provide adequate fitting to empirical results from wireless channel measurements in both indoor and outdoor communication scenarios. Furthermore, it is capable of accounting for propagation in urban environments in cases, where Rayleigh distribution is
inadequate, as well as in modelling path-loss in, e.g., Digital Enhanced Cordless Telecommunications (DECT) system at 1.89 GHz \cite{J:Weibull, J:Karagiannidis_1, J:Matolak, C:Stefanovic, J:Dias, B:Alouini, J:Fraidenraich}. Its flexibility is also evident from the fact that it includes as special cases the well-known Rayleigh and exponential distributions for $a = 2$ and $a = 1$, respectively.

However, in spite of the undoubted usefulness of Weibull fading model, no studies related to the detection of unknown signals over such fading conditions have been reported in the open technical literature. Motivated by this, the present work is devoted to the analytical performance evaluation of energy detection over narrowband Weibull fading channels. Specifically, a novel analytic expression is firstly derived for the average probability of detection of the energy detector. This expression is rather simple both algebraically and numerically and is subsequently used to analyze the performance of different fading conditions.  As expected, the performance of the energy detector is highly dependent upon the value and variations of the involved fading parameters. This allows the quantification of the effect of fading in the system performance in various communication scenarios which can enable the determination of the required power levels for ensuring robust and accurate performance of energy detectors along with sufficient benefits, when possible, with respect to energy efficiency.

The remainder of this paper is organized as follows: The system and channel model are described in Section II. The average detection probability of the energy detector over Weibull fading channels is analyzed in Section III and numerical results for each communication scenarios and discussions are provided in Section IV. Finally,   closing remarks are given in Section V.


\section{System and Channel Model}

\subsection{Energy Detection}

The received signal waveform in narrowband energy detection follows a binary hypothesis that can be expressed as \cite[eq. (1)]{C:Beaulieu},

\begin{equation} \label{Test_1} 
r(t) =
\begin{cases}
n(t) \,\,\,\, \qquad \,\,\,\, \qquad \,\,\, \,\,\,\,\,\,\,\,\,\,\,\,\,\,\,\,\,\,:H_{0} \\ 
hs(t) + n(t) \, \qquad \, \,\,\,\,\,\,\,\,\,\,\,\,\,\,\,\,\,\,\,:H_{1} 
\end{cases}
\end{equation}
where $s(t)$ is an unknown deterministic signal whereas $h$ denotes the complex gain of the channel coefficient and $n(t)$ is an additive white Gaussian noise (AWGN) process. The samples of $n(t)$ are assumed to be zero-mean Gaussian random variables with variance $N_{0}W$ with $W$ and $N_{0}$ denoting the  single-sided signal bandwidth and a single-sided noise power spectral density, respectively  \cite{C:Beaulieu}. The hypotheses $H_{0}$ and $H_{1}$ refer to the cases that a signal is absent or present, respectively.  The received signal is subject to filtering, squaring and integrating over the time interval $T$ which is given by \cite[eq. (2)]{J:Alouini} as 

\begin{equation}
y \triangleq \frac{2}{N_{0}} \int_{0}^{T} \mid r(t)\mid ^{2} dt
\end{equation}

The output of the integrator is a measure of the energy of the received waveform which constitutes a test statistic that determines whether the received energy measure corresponds only to the energy of noise ($H_{0}$) or to the energy of both the unknown deterministic signal and noise ($H_{1}$). By denoting the observation time bandwidth product as $u = TW$,  the test statistic follows the central chi-square distribution with $2u$ degrees of freedom under the $H_{0}$ hypothesis and the non central chi-square distribution with $2u$ degrees of freedom under the $H_{1}$ hypothesis \cite{J:Urkowitz}. To this effect, the corresponding probability density function (PDF) in the presence of AWGN is expressed according to \cite[eq. (3)]{J:Alouini}, namely, 
 
\begin{equation} \label{Test_3} 
p_{Y}(y) = 
\begin{cases}
\frac{1}{2^{u}\Gamma(u)}y^{u-1}e^{-\frac{y}{2}} \,\, \qquad \,\, \qquad \,\, \qquad \,\,\,\,\, \,\,\,\,:H_{0} \\ 
\frac{1}{2}\left(\frac{y}{2\gamma} \right)^{\frac{u-1}{2}} e^{-\frac{y + 2\gamma}{2}}I_{u-1}\left(\sqrt{2y\gamma} \right) \, \, \,\,\,\,\,\,\,\,:H_{1}
\end{cases}
\end{equation}
where

\begin{equation}
\gamma \triangleq   |h| ^{2} \frac{E_{s}}{N_{0}}
\end{equation}
is the corresponding instantaneous SNR and $E_{s}$ denotes the signal energy. Furthermore, 
 
\begin{equation} \label{gamma_function}
\Gamma\left(a \right) \triangleq \int_{0}^{\infty} t^{a-1}e^{-t}dt,
\end{equation}
stands for the Euler's gamma function and
 
\begin{equation}
I_{n}\left(x\right) \triangleq \frac{1}{\pi} \int_{0}^{\pi} \cos(n \theta) e^{x \cos(\theta) } d\theta, 
\end{equation}
denotes the modified Bessel function of the first kind \cite{B:Abramowitz}.

As already mentioned, an energy detector is largely characterized by a predefined energy threshold, $\lambda$. This threshold is particularly critical in the decision process and is promptly associated to three measures that overall evaluate the performance of the detector: i) the probability of false alarm, $P_{f}= {\rm Pr}(y> \lambda \mid H_{0})$;  ii) the probability of detection, $P_{d}= {\rm Pr}(y> \lambda \mid H_{1})$ and iii) the probability of missed detection, $P_{m} = 1 - P_{d}$. The first two measures are deduced by integrating \eqref{Test_3} over the interval between the energy threshold to infinity, $ \left\lbrace   \lambda, \, \infty  \right\rbrace $, yielding \cite{J:Alouini}, 
 
\begin{equation} \label{Pf_1} 
P_{f} = \frac{\Gamma \left(u,\frac{\lambda}{2}\right)}{\Gamma(u)}, 
\end{equation}
\noindent 
and
 
\begin{equation}\label{Pd_1} 
P_{d} = Q_{u}(\sqrt{2 \gamma},\sqrt{\lambda}),
\end{equation}
\noindent 
where,
 
\begin{equation}
\Gamma\left(a, x\right) \triangleq \int_{x}^{\infty} t^{a-1}e^{-t}dt,
\end{equation}
denotes the upper incomplete gamma function \cite{B:Abramowitz} and 
 
\begin{equation}
Q_{m}(a,b) \triangleq \frac{1}{a^{m-1}} \int_{b}^{\infty} x^{m} e^{-\frac{x^{2} + a^{2}}{2}} I_{m - 1} (ax)dx, 
\end{equation}
is the generalized Marcum $Q{-}$function \cite{J:Marcum}.

\subsection{The Weibull Fading Distribution}

As already mentioned, the Weibull  fading model has been shown to represent effectively the small-scale variations of a signal in non-line-of-sight (NLOS) communications in multipath environments. Physically, this fading model  accounts for propagation in non-homogeneous environments and the resulting envelope is
obtained as a nonlinear function of the modulus of the sum of the involved multipath components \cite{C:Stefanovic}.  The envelope PDF of Weibull distribution is given by \cite[eq. (2.27)]{B:Alouini}, namely, 
 
\begin{equation} \label{envelope_pdf_weibull} 
p_{_{R}}(r) =  a \left[\frac{\Gamma \left(1 + \frac{2}{a} \right)}{\Omega} \right]^{\frac{a}{2}} r^{a-1} e^{- \left[ \frac{r^{2}}{\Omega} \Gamma\left(1 + \frac{2}{a} \right) \right]^{\frac{a}{2}}}, 
\end{equation}
where $a$ denotes the fading parameter that is selected to yield the best fit to measurement results and $\Omega = \overline{R^{a}} = E[R^{a}]$ is the  mean power of the desired signal \cite{B:Alouini}. The corresponding PDF  of the instantaneous SNR per symbol $\gamma$ is given by \cite[eq. (2.29)]{B:Alouini}, namely, 
 
\begin{equation} \label{power_pdf_weibull} 
p_{\gamma}(\gamma) =  \frac{a}{2} \left[\frac{\Gamma \left(1 + \frac{2}{a} \right)}{\overline{\gamma}} \right]^{\frac{a}{2}} \gamma^{\frac{a}{2} - 1} e^{- \left[ \frac{\gamma}{\overline{\gamma}} \Gamma\left(1 + \frac{2}{a} \right) \right]^{\frac{a}{2}}},  
\end{equation}
where $\overline{\gamma}$ represents the average SNR per symbol. To this effect, the corresponding envelope and power CDFs are expressed as,  
 
\begin{equation} \label{envelope_cdf_weibull} 
P_{_{R}}(r) =  1 - e^{- \left[ \frac{r^{2}}{\Omega} \Gamma\left(1 + \frac{2}{a} \right) \right]^{\frac{a}{2}}},
\end{equation}
and
 
\begin{equation} \label{power_cdf_weibull} 
P_{\gamma}(\gamma) =  1 -  e^{- \left[ \frac{\gamma}{\overline{\gamma}} \Gamma\left(1 + \frac{2}{a} \right) \right]^{\frac{a}{2}}}, 
\end{equation}
respectively \cite{B:Alouini}. The $n^{th}$ statistical moment is expressed as,
 
\begin{equation} \label{moments}
E[R^{n}] = \Gamma \left[ 1 + \frac{n}{a} \right], 
\end{equation} 
whereas the corresponding MGF is given by \cite[eq. (2.35)]{B:Alouini}, 
 
\begin{equation} \label{MGF}
M_{R}(s)  = \left[ \frac{\Gamma \left(1 + \frac{2}{a} \right)}{\Omega} \right]^{\frac{a}{2}} \frac{(-1)^{a} a^{a + \frac{1}{2}} }{  (2 \pi)^{\frac{a - 1}{2}} s^{a}}  G^{c,1}_{1,c} \left\lbrace (-1)^{a} \left[ \frac{\Omega}{\Gamma\left(1 + \frac{2}{a} \right) }    \right]^{\frac{a}{2}} \frac{s^{a}}{a^{a}} \Bigg| ^{\, \quad \, \quad \, 1}_{1, 1 + \frac{1}{a}, \cdots, \frac{a - 1}{a}} \right\rbrace, 
\end{equation} 
where $G^{.,.}_{.,.} (.)$ denotes the Meijer $G{-}$function \cite{B:Tables}.   


\section{Average Detection Probability over Weibull Fading Channels}

It is recalled that \eqref{Pf_1} and \eqref{Pd_1} account for the case of AWGN channels. For communication scenarios over fading channels the average probability of detection is obtained by averaging \eqref{Pd_1} over the corresponding SNR fading statistics, namely,
 
\begin{equation} \label{Av_Pd_1} 
\overline{P_{d}} = \int_{0}^{\infty }Q_{u}(\sqrt{2 \gamma},\sqrt{\lambda})p_{\gamma}(\gamma) d\gamma .
\end{equation}
Based on this, the average probability of detection in the case of Weibull fading channels can be obtained by averaging \eqref{Pd_1} over the statistics of \eqref{power_pdf_weibull}. To this effect, by substituting \eqref{power_pdf_weibull} in \eqref{Av_Pd_1} it immediately follows that, 
 
\begin{equation} \label{Av_Pd_2}  
\overline{P_{d}} =  \frac{a}{2} \left[\frac{\Gamma \left(1 + \frac{2}{a} \right)}{\overline{\gamma}} \right]^{\frac{a}{2}}   \int_{0}^{\infty }  \frac{ \gamma^{\frac{a}{2} - 1}  Q_{u}\left(\sqrt{2 \gamma},\sqrt{\lambda}\right)}{e^{\left[ \frac{\gamma}{\overline{\gamma}} \Gamma\left(1 + \frac{2}{a} \right) \right]^{\frac{a}{2}}}}       d\gamma.  
\end{equation}
Evidently, an analytic expression for the $\overline{P_{d}} $ in Weibull fading conditions is subject to analytical solution of the integral in \eqref{Av_Pd_2} which is not available in tabulated form in the open technical literature.  To this end, we firstly let $x = \sqrt{2\gamma}$ and thus: $\gamma = x^{2}{/}2$ and $dx{/}d\gamma = 1{/}\sqrt{2\gamma}$. By substituting in \eqref{Av_Pd_2} and after some basic algebraic manipulations one obtains, 
 
\begin{equation} \label{Av_Pd_3}  
\overline{P_{d}} =  \frac{\mathcal{A}  a}{(2 \overline{\gamma})^{\frac{a}{2}}  } \underbrace{ \int_{0}^{\infty }    x^{a- 1}  Q_{u}\left(x, \sqrt{\lambda}\right)   e^{- \frac{\mathcal{A} x^{a}}{(2 \overline{\gamma})^{ \frac{a}{2} }}   }    dx }_{\mathcal{I}_{1}}, 
\end{equation}
where,
 
\begin{equation} \label{constant_1}
\mathcal{A} = \left[ \Gamma \left( 1 + \frac{2}{a} \right) \right] ^{ \frac{a}{2} }. 
\end{equation}
By integrating $\mathcal{I}_{1}$ by part it follows that, 
 
\begin{equation} \label{Integral_1}
\mathcal{I}_{1}   =   \lim_{x \rightarrow 0} \frac{(2 \overline{\gamma})^{\frac{a}{2}} Q_{u}(x, \sqrt{\lambda})  }{a \, \mathcal{A} \, e^{ \frac{\mathcal{A} x^{a} }{(2 \overline{\gamma})^{ a{/}2 }} }  } - \lim_{x \rightarrow \infty} \frac{(2 \overline{\gamma})^{\frac{a}{2}} Q_{u}(x, \sqrt{\lambda})  }{a \, \mathcal{A} \, e^{ \frac{\mathcal{A} x^{a} }{(2 \overline{\gamma})^{ a{/}2 }} }  }   +  \frac{(2 \overline{\gamma})^{\frac{a}{2}} }{a \, \mathcal{A} \,  } \int_{0}^{\infty}   e^{ - \frac{\mathcal{A} x^{a} }{(2 \overline{\gamma})^{ a{/}2 }} } \left[  \frac{d}{dx} Q_{u}(x, \sqrt{\lambda}) \right]  dx.    
\end{equation}
By recalling that $Q_{u}(\infty, b) \triangleq 1 $ and based on \cite[eq. (2)]{J:Nuttall_2} and \cite[eq. (10)]{J:Nuttall_2}, it immediately follows that,  
 
\begin{equation} \label{Integral_2}
Q_{u}(0, \sqrt{\lambda}) =  e^{- \frac{\lambda}{2} } \sum_{l = 0}^{u - 1} \frac{\lambda^{l}}{l! 2^{l}},  
\end{equation}
and
 
\begin{equation} \label{Integral_3}
\frac{d}{dx}  Q_{u}(x,\sqrt{\lambda}) = \frac{\lambda^{\frac{u}{2}}}{x^{u-1}} e^{ - \frac{x^{2} + \lambda}{2} } I_{u}(x \sqrt{\lambda}), 
\end{equation}
respectively. Thus, by substituting \eqref{Integral_2} and \eqref{Integral_3} in \eqref{Integral_1} yields,
 
\begin{equation} \label{Integral_4}
\mathcal{I}_{1} = \sum_{l = 0}^{u-1} \frac{\lambda^{l} e^{- \frac{\lambda}{2} }}{l! 2^{l} } +   \frac{ \lambda^{\frac{u}{2}} (2 \overline{\gamma})^{\frac{a}{2}} }{a \, \mathcal{A} \, e^{ \frac{\lambda}{2} }  }  \underbrace{ \int_{0}^{\infty}    \frac{ x^{1 - u} I_{u}(x \sqrt{\lambda}) }{ e^{ \frac{\mathcal{A} x^{a} }{(2 \overline{\gamma})^{ a{/}2 }} }  e^{ \frac{x^{2} }{2} } }    dx}_{\mathcal{I}_{2}}. 
\end{equation}
By expanding the exponential term with the arbitrary power, the $\mathcal{I}_{2}$ term can be equivalently  written as follows,  
 
\begin{equation} \label{Integral_5}
\mathcal{I}_{2} = \sum_{l = 0}^{\infty} \frac{ (-1)^{l} \mathcal{A}^{l} }{l! (2 \overline{\gamma} )^{ a l {/}2 } }  \underbrace{  \int_{0}^{\infty}     x^{1 - u + la} e^{ - \frac{x^{2} }{2} }  I_{u}(x \sqrt{\lambda})  \,   dx }_{\mathcal{I}_{3}}.  
\end{equation}
Importantly, the $\mathcal{I}_{3}$ integral can be expressed in closed-form according to \cite[eq. (6.643.2)]{B:Tables}. To this effect, by making the necessary change of variables and substituting in \eqref{Integral_5} and subsequently in \eqref{Integral_4} and \eqref{Av_Pd_3}, the following analytic expression   is deduced for the average probability of detection,   
 
\begin{equation} \label{Final}
\overline{P_{d}}  = \sum_{l = 0}^{u-1} \frac{\lambda^{l} e^{- \frac{\lambda}{2} } }{l! 2^{l}}    + \sum_{l = 0}^{ \infty} \frac{(-1)^{l} \mathcal{A}^{l} \lambda^{u}  }{l! u! 2^{u} \overline{\gamma}^{\frac{l a }{2}} e^{  \frac{\lambda}{2}}  } \Gamma\left( \frac{l a }{2} + 1  \right) \, _{1}F_{1}\left(\frac{l a }{2} + 1, u+1, \frac{\lambda}{2}  \right),  
\end{equation}
where,
 
\begin{equation} \label{hypergeometric}
\, _{1}F_{1} (a, b, x) \triangleq \sum_{l = 0}^{\infty}  \frac{(a)_{l}}{(b)_{l}} \frac{x^{l}}{l!}, 
\end{equation}
is the Kummer's confluent hypergeometric function and 

\begin{equation}
(a)_{n} \triangleq  \frac{\Gamma(a + n)}{\Gamma(a)}
\end{equation}
denotes the Pochhammer symbol  \cite{B:Abramowitz}. It is noted that the series converges rapidly and only few terms are required for achieving acceptably low levels of  truncation error.  This is particularly important since spectrum sensing applications   typically require accuracy of two  decimal digits.

To the best of the Authors' knowledge, equation \eqref{Final} has not been  reported in the open technical literature.


\section{Numerical Results }

Having derived a novel analytic expression for  the average probability of detection,  this section is devoted to  the analysis
\begin{figure}[h!]
\centering
\includegraphics[ width=12cm,height=9cm]{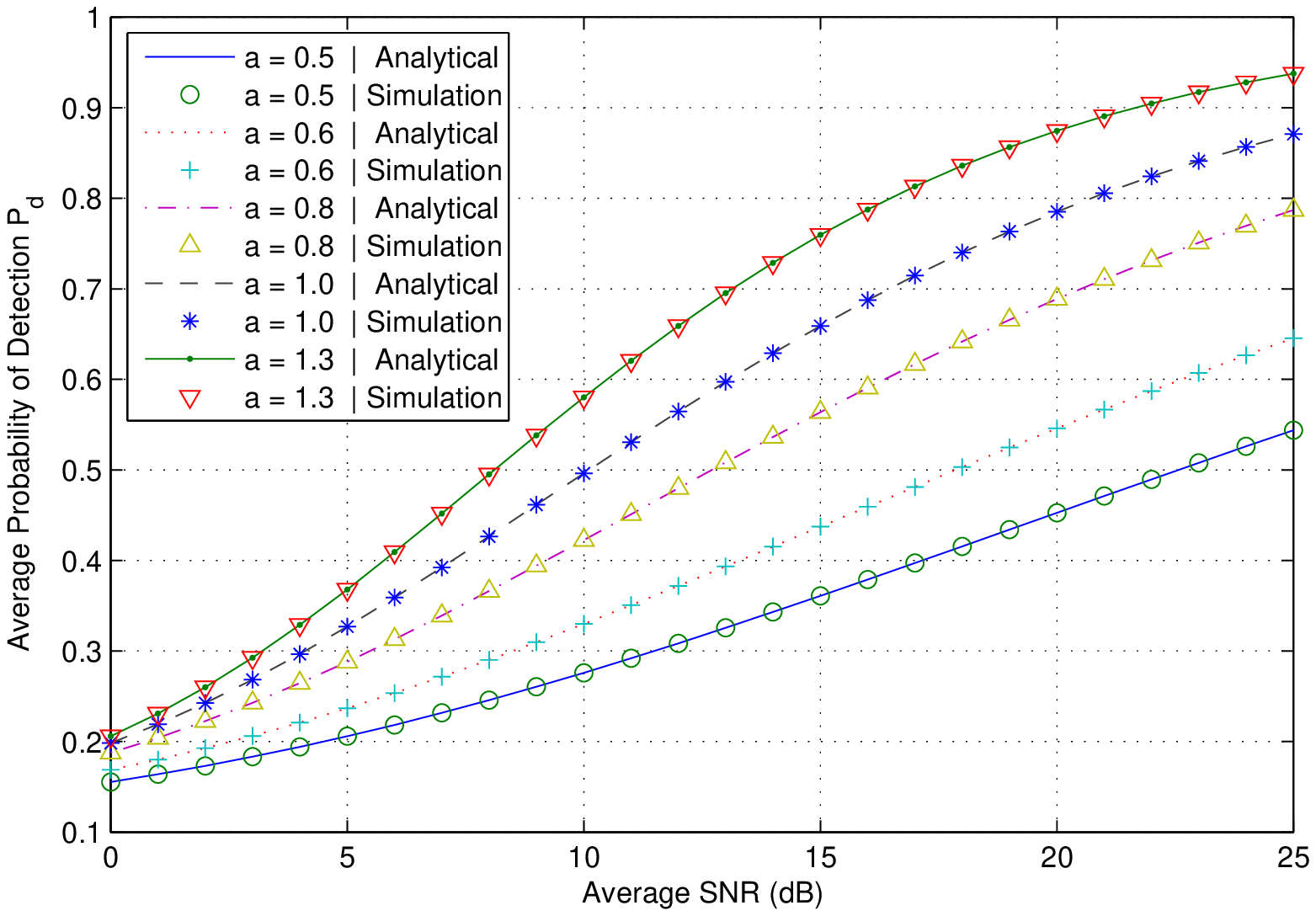} 
\caption{ $\overline{P}_{d}$ vs $\overline{\gamma}$ for i.i.d Weibull fading with $P_{f} = 0.1$, $u = 5$ and different values for fading severity $a$.} 
\end{figure}
 of the behaviour of energy detection over Weibull fading channels. The corresponding performance is evaluated for different scenarios of interest through both $\overline{P}_{d}$ versus $\overline{\gamma}$ curves and complementary receiver operating characteristics (ROC) curves ($P_{m}$ versus $P_{f}$). To this end, 
\begin{figure}[h!]
\centering
\includegraphics[ width=12cm,height=9cm]{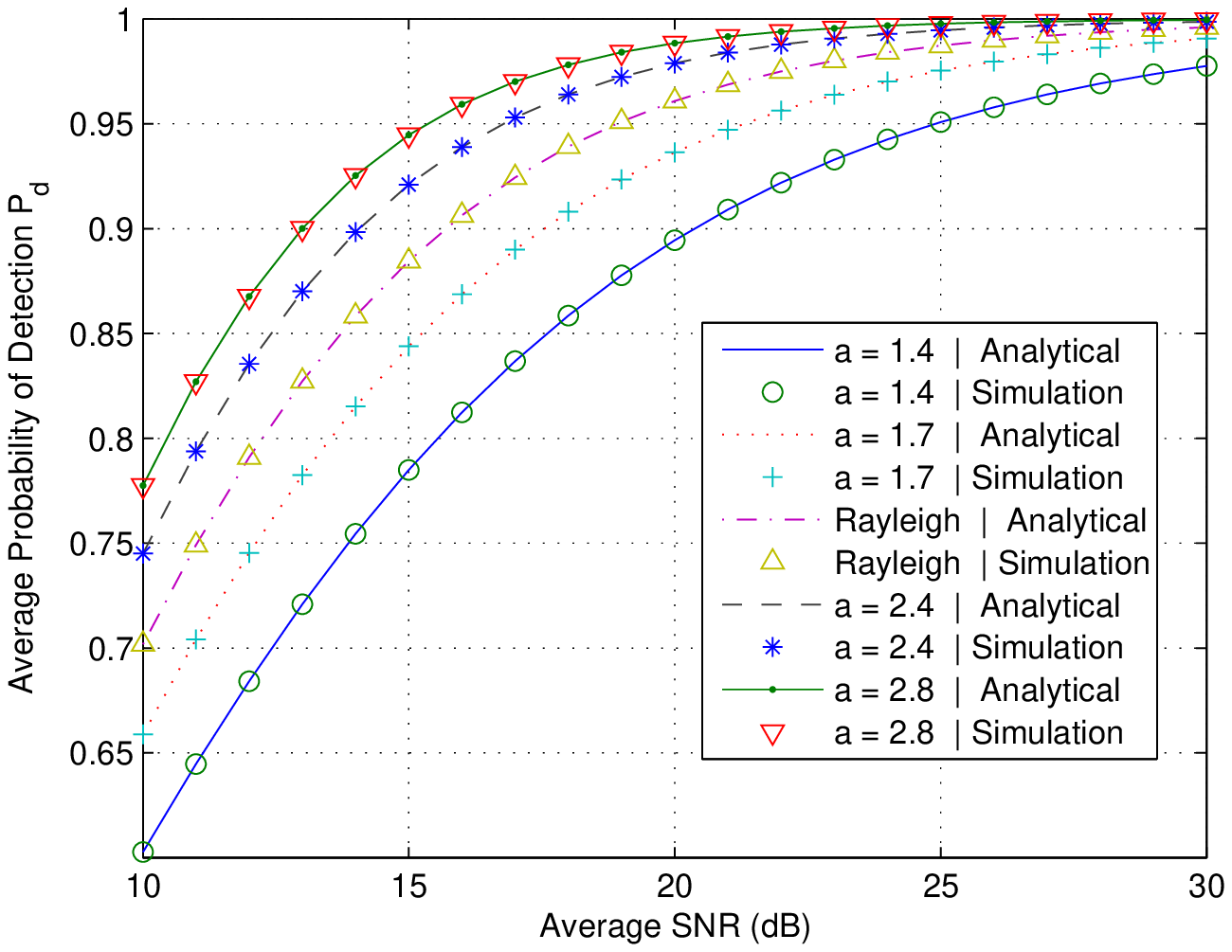} 
\caption{ $\overline{P}_{d}$ vs $\overline{\gamma}$ for i.i.d Weibull fading with $P_{f} = 0.1$, $u = 5$ and different values for fading severity $a$.} 
\end{figure}
 Fig. $1$ demonstrates $\overline{P}_{d}$ vs $\overline{\gamma}$ curves  for different values of fading severity $a$ with $P_{f} = 0.1$ and $u=5$. One can clearly notice   that the behaviour of the energy detector is very sensitive as even at slight variations of $a$, the value of $\overline{P}_{d}$ changes rapidly. As expected, the performance of the detector is directly proportional to the value of the severity index $a$ and the average SNR $\overline{\gamma}$. For small values of $a$ and $\overline{\gamma}$ the performance of the detector for $P_{f} = 0.1$ is rather low and thus, increasing  the time-bandwidth product $u$ is essential. 
 \begin{figure}[h!]
 \centering
\includegraphics[ width=12cm,height=9cm]{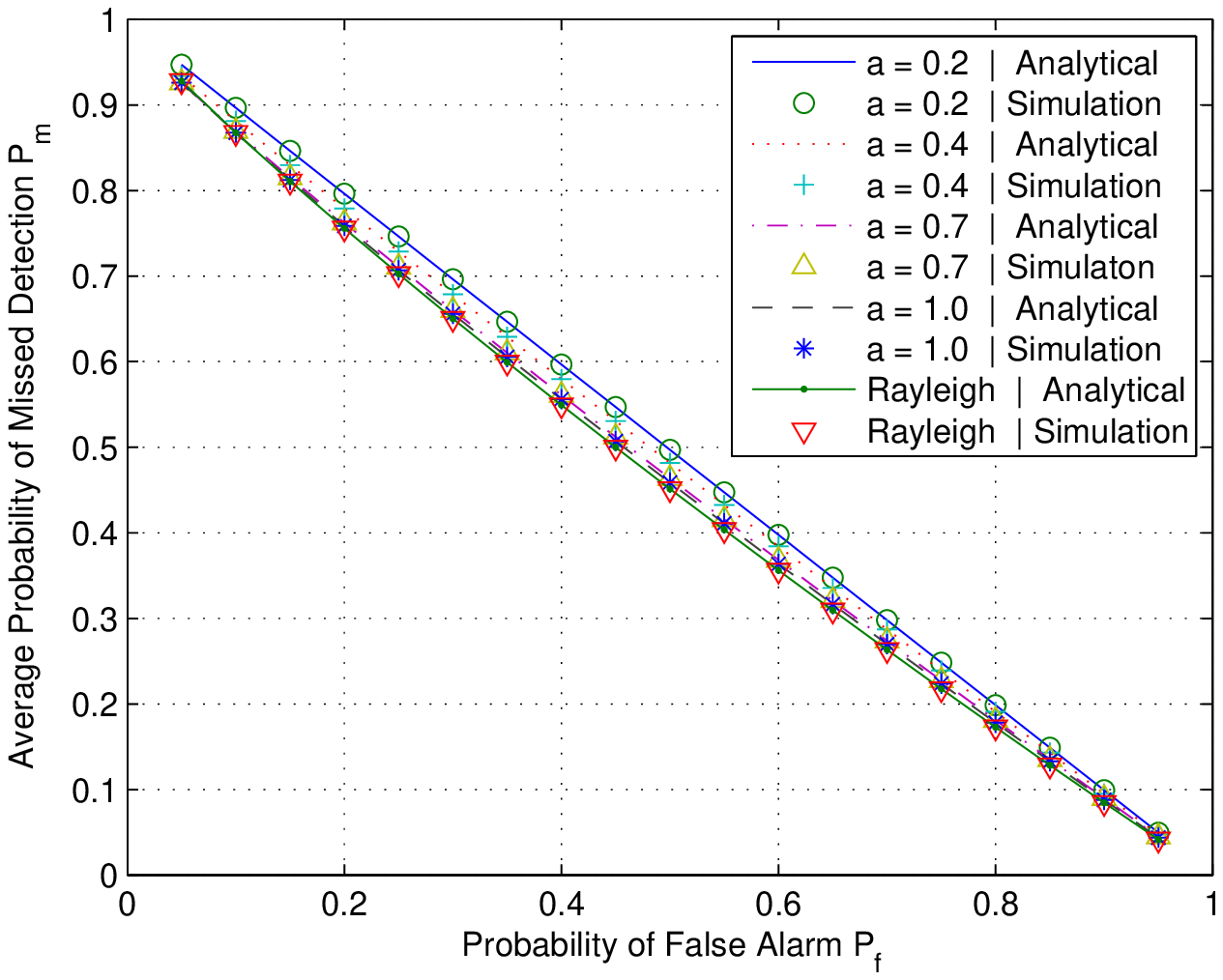} 
\caption{ Complementary ROC curves for   $u=5$, $\overline{\gamma}  = -5\,$dB and  different values of $a$. }  
\end{figure}
Likewise, Fig. $2$ depicts the behaviour of $\overline{P}_{d}$ vs $\overline{\gamma}$  for higher values of $a$. The performance of the detector increases significantly, particularly in the post-Rayleigh area i.e. $a > 2$. Furthermore, the overall flexibility of the Weibull model is evident since it can account for a wide range of severity conditions that have a significant effect on the corresponding performance.

\begin{figure}[h!]
\centering
\includegraphics[ width=12cm,height=9cm]{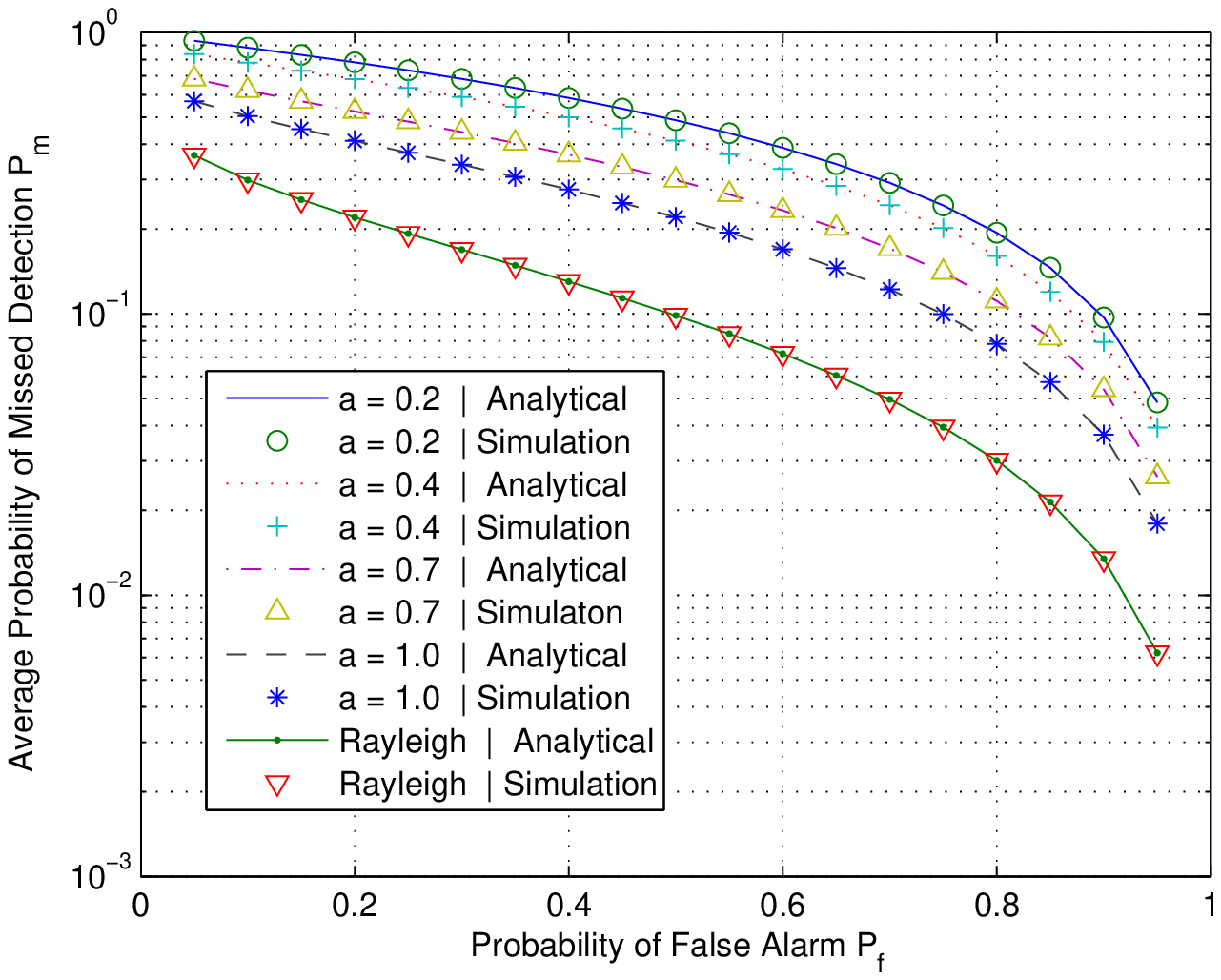} 
\caption{ Complementary ROC curves for   $u=5$, $\overline{\gamma}  = 10\,$dB and  different values of $a$. }  
\end{figure}

\begin{figure}[h!]
\centering
\includegraphics[ width=12cm,height=9cm]{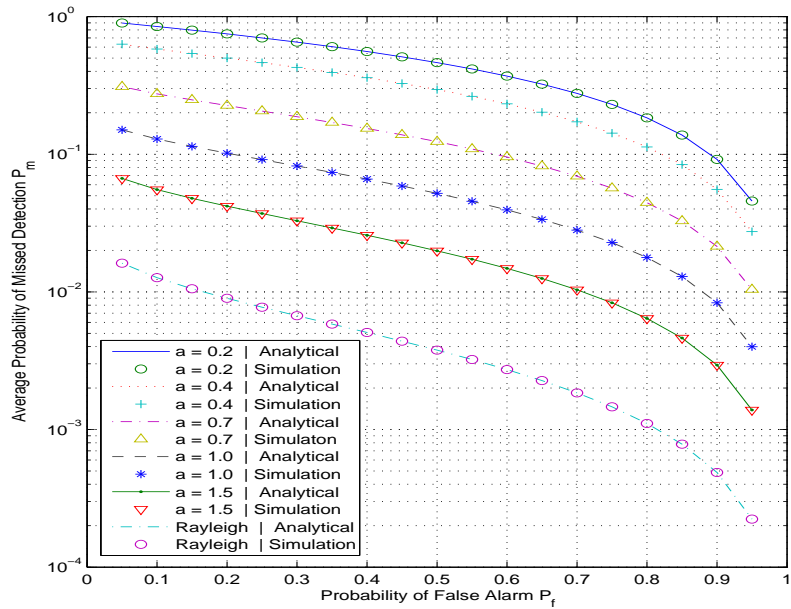} 
\caption{ Complementary ROC curves for   $u=5$, $\overline{\gamma}  = 25\,$dB and  different values of $a$. }  
\end{figure}
Figs. $3{-}5$  illustrate complementary receiver-operating-characteristics (ROC) curves $P_{m} $ vs $P_{f}$ for $u = 5$ at: $i)$ low SNR ($\overline{\gamma} = -5$dB); $ii)$ moderate SNR ($\overline{\gamma} = 10$dB); $iii)$ high SNR  ($\overline{\gamma} = 25$dB) and different severe fading conditions. It is clearly shown that the effect of severe fading on the system performance is stronger as the average SNR increases. For example, for $P_{f} = 0.2$ and $a = 1.0$, the corresponding value for the probability of missed detection is $P_{m} = 0.78$, $P_{m} = 0.41$ and $P_{m} = 0.1$, respectively. However, it generally appears that the detector can perform  adequately even at significantly more severe fading conditions that Rayleigh fading. This also verifies the usefulness of Weibull distribution as an adequate model for accounting for the multipath fading effect.

\section{Conclusion}

This work analyzed the performance of energy detection over Weibull fading channels. A novel analytic expression was derived for the average probability of detection and was shown that the overall performance of the detector is largely affected by the value of the involved parameters.  This is evident  by the fact that even slight variations of the fading parameter results to substantial variation of the corresponding probability of detection.   As a result, the offered results are useful in quantifying the effect of fading in energy detection spectrum sensing which can contribute towards improved and    energy efficient cognitive radio wireless communication deployments.

{}

\end{document}